# 0.5 keV soft X-ray attosecond continua

S. M. Teichmann[1], F. Silva[1], S. L. Cousin[1], M. Hemmer[1], J. Biegert[1,2]

[1]ICFO-Institut de Ciencies Fotoniques, The Barcelona Institute of Science and Technology,
08860 Castelldefels (Barcelona), Spain

[2]ICREA-Institució Catalana de Recerca i Estudis Avançats, 08010 Barcelona, Spain

e-mail: jens.biegert@icfo.eu

**Attosecond light pulses in the extreme ultraviolet have drawn a great deal of attention due to their ability to interrogate electronic dynamics in real time. Nevertheless, to follow charge dynamics and excitations in materials, element selectivity is a prerequisite, which demands such pulses in the soft X-ray region, above 200 eV, to simultaneously cover several fundamental absorption edges of the constituents of the materials. Here, we experimentally demonstrate the exploitation of a transient phase matching regime to generate carrier envelope controlled soft X-ray supercontinua with pulse energies up to 2.9 ± 0.1 pJ and a flux of (7.3 ± 0.1)x$10^7$ photons/s across the entire water window and attosecond pulses with 13 as transform limit. Our results herald attosecond science at the fundamental absorption edges of matter by bridging the gap between ultrafast temporal resolution and element specific probing.**

The availability of waveform controlled pulsed coherent radiation covering the entire water window represents a current frontier in the development of ultrafast soft X-ray sources since this spectral range contains the K-shell absorption edges of carbon (284 eV), nitrogen (410 eV), and oxygen (543 eV) as well as the L-shell absorption edges of calcium (341 eV), scandium (395 eV), titanium (456 eV) and vanadium (511 eV). Controlled coherent radiation consisting of isolated attosecond pulses in this photon energy region would herald a new era of attosecond physics since it would combine its unprecedented temporal resolution with the ability to localize the observed dynamics at a fundamental absorption edge, i.e. at a specific atom or site inside a material and follow it to another atom or site. Such attosecond soft X-ray spectroscopy would offer site-specific probes for observing electron correlation [1-4] and many-body effects of core-excited atoms [5,6] or attosecond electron transfer [7] in photo- and electro-chemical processes of organic solar cells and molecular electronics. This would permit investigating semiconductor surfaces where charge transfer occurs on a timescale shorter than 3 fs [8], and to study insulator-to-metal transitions at the oxygen edge [9]. With water window covering soft X-ray radiation, coherent imaging measurements such as transmissive, reflective and ptychographic diffractive imaging can be envisaged that offer the possibility to resolve structural dynamics of biomolecules at high resolution and fast timescales.

Attosecond light pulses [10] are generated through coherent frequency upconversion of intense laser pulses via high harmonic generation (HHG), a process that relies on sub-cycle electron recollision in the laser field and recombinative light emission. These dynamics occur during a fraction of the driving laser electric field period and naturally lead to the emission of attosecond or femtosecond bursts of radiation. Contributions from each wave-cycle within a long laser pulse result in the generation of trains of such pulses. On the other hand, limiting the recombination to a single wave-cycle [11] by using a carrier to envelope phase (CEP) stable sub-2-cycle laser field results in single attosecond pulse emission [11-14]. High photon energies are achieved through ponderomotive scaling by driving the HHG process with long wavelength lasers [15-20]. With such approach, first CEP dependent HHG spectra were demonstrated whose cutoff reached 300 eV [18,19]. We have recently validated the viability of such a HHG-based source for soft X-ray absorption fine structure (XAFS) measurement at the carbon K-edge from a solid material [19], and we have shown that isolated attosecond structures with duration below 355 as can be produced at 300 eV [20].

Here, we address the to-date unsolved challenge of phase matching CEP controlled HHG with coverage from 200 eV across the entire water window up to the oxygen K-shell edge at 543 eV. Such an achievement is markedly different from coverage of a single absorption edge at 300 eV since only the simultaneous measurement at several absorption edges will permit following an excitation or charge density across the various constituents of a material. For instance, coverage of the chlorine, carbon and nitrogen edges simultaneously, will permit following the formation of an exciton at the chlorine (205 eV) site of the antenna complex of an organic solar cell and its diffusion towards the charge separating interface across the cell's hydrocarbon backbone. We demonstrate experimentally, and investigate theoretically, the regime of high-flux phase matching of soft X-ray continua with CEP control and coverage of the entire water window. Our source now enables interrogation of the most fundamental absorption edges within the water window simultaneously and with attosecond temporal resolution.

**Phase matched soft-X-ray emission covering the entire water window.** A schematic of our setup is shown in Fig 1. Carrier-envelope-phase (CEP) stable (see Supplementary Information for details), 1.9-cycle pulses at 1850 nm wavelength at a repetition rate of 1 kHz are focused with a 100 mm focal length spherical mirror to a peak intensity of 0.5 PW/cm$^2$ to generate high harmonic radiation. The HHG target consists of a tube with 1.5 mm outer diameter with 0.3 mm entrance and exit holes through which the laser beam is focused; see the Methods section for further details. A differential pumping scheme allows us to sustain 6 bar of backing pressure with helium while maintaining an ambient pressure of 20 mbar in the generation chamber. Measurements up to 12.5 bar in helium are possible but cannot be sustained for a long time with our present vacuum system. High harmonic radiation is spectrally resolved with a homebuilt spectrograph consisting of a motorized slit, a flat-field imaging reflective grating (Hitachi, 2400 lines/mm) and a cooled, back-illuminated soft X-ray charge-coupled device camera (PIXIS-XO-2048B, Princeton Instruments). For a first study, we varied the target backing pressure for neon (Fig. 1A) which gave an optimum yield at 3.5 bar and cutoff at 350 eV, amply covering the carbon K-shell edge at 284 eV. The strongest emission in helium coincided with the highest cutoff at 550 eV, which was obtained at the maximum backing pressure of 12.5 bar (Fig. 1B). This indicates that higher pressures need to be achieved than available in our setup to assess the exact pressure dependence and possible even stronger harmonic signal. Having determined the optimum pressure conditions, we opted to use helium at the maximum sustainable pressure of 6 bar. Figure 1C displays the measured sub-2-cycle driven soft X-ray continuum, on a linear scale, covering the entire water window up to the oxygen edge at 543 eV. Clearly resolved are the soft X-ray absorption K-shell edges resulting from the insertion of 200 nm thin films of carbon (Fig. 1D) and titanium (Fig. 1E) which are visible at 284 eV and 456 eV respectively when integrating for 2 min. An important parameter in HHG is the position of the generation target relative to the focal position as it is known to sensitively influence phase matching and the contribution of the various quantum paths to harmonic emission [21]. Here, we observe a different behavior to the known trends for the conditions of maximum yield and highest cutoff for both target gases that permit reaching the water window, i.e. for neon and helium. We do not find any significant variation of the cutoff yield when scanning the HHG target within the Rayleigh length, i.e. for our conditions within 1 mm on either side of the focal plane. Scanning slightly beyond the Rayleigh length, we observe an immediate loss of signal. This is markedly different from the known smoother transition, and slow decrease of signal, for low pressure phase matching and 800 nm wavelength [21-23] and will be discussed later.

**Soft X-ray yield and CEP controlled spectra.** The photon yield and spectral stability of a source are essential for its applications. Having observed the peculiar phase matching dependence on the focusing and target conditions, we investigate its influence on the usability of such HHG soft X-ray source. Such investigation is important since the photon yield needs to be sufficient to permit discrimination of the measurement signal from the noise, and the stability and reproducibility of the spectrum are essential for spectroscopic applications. Hence, we investigate the CEP control and reproducibility of the soft X-ray spectra. Figure 2 shows the spectra measured for a range of CEP values in steps of 90 mrad and for 30 second integration. The clear discrimination between the spectra (Fig. 2A,C) and their reproducibility for an offset of π rad (Fig. 2B,D) demonstrates the stability and utility of our soft X-ray source. The importance of CEP control for spectroscopic measurements with ultrafast time resolution is underlined in Fig. 2B and D which exhibit a strong variation of the spectral envelope, and shift of the cutoff, for two different CEP values in neon and helium. The observed spectral changes, and cutoff shifts of up to 10%, of the cutoff energy would have detrimental influence on the utility of harmonic sources for spectroscopic measurements without the demonstrated control and stability. Corresponding photon fluxes of the source at 284 eV are $(2.8 \pm 0.1) \times 10^7$ photons/s/10% bandwidth (BW) in neon and $(1.8 \pm 0.1) \times 10^6$ photons/s/10% BW in helium, resulting in pulse energies, defined only in the water window, between the carbon K-shell edge and the cutoff, of 2.9 ± 0.1 pJ in neon and 0.9 ± 0.2 pJ in helium. Clearly, these energies are sufficient for soft X-ray absorption spectroscopy as shown by the measurements in Fig. 1D,E.

It is interesting to contrast our results with previous work. Our water window flux of $(7.3 \pm 0.1) \times 10^7$ photons/s from neon is similar to the previous reported record value of $6 \times 10^7$ photons/s which was however achieved for a non-CEP-stable source and a multi-cycle (40 fs, or >6 cycles) pulse at 2 μm [24] – note that those driving laser pulses did not result in reproducible isolated attosecond pulse emission. While the achieved general parameters may look similar at first glance, the decisive difference is that we have achieved unprecedented flux for a sub-2-cycle driven CEP controlled soft X-ray continuum which corresponds to a reproducible single attosecond pulse [25]. The measured soft X-ray continua support extremely short durations of 18 as and 13 as in neon and helium, respectively. Moreover, driving soft X-ray continua and attosecond generation with a sub-2-cycle pulse is indeed preferable as we have recently demonstrated that a high level of ionization, such as for the conditions of high pressure phase matching and long wavelengths, has an adverse effect on spectral stability and it could prevent reproducible attosecond pulse generation altogether [26]. Alternative approaches to generate spectral continua, and isolated attosecond pulses, from multi-cycle driving pulses rely on some form of temporal confinement of recollision such that it effectively occurs only once [11, 27, 28]. These implementations work very well for 800 nm driven HHG, but they are severely limited for long wavelength driven HHG due to the required much higher phase matching pressures and the consequently high ionization levels. This limitation is evident from Ref. [29] in which ionization gating was implemented for long wavelength (2 μm) driven multi-cycle HHG. The conditions to achieve a spectral continuum dictated a much lower pressure of 600 torr (0.7 bar) which consequently limited the phase matchable cutoff to 175 eV - far below the water window. This result points at possible severe limitations in implementing the well established gating concepts to achieve attosecond radiation in the water window soft X-ray range. Our implementation circumvents such problems and it is therefore interesting to investigate the conditions of phase matching.

**Transient phase matching of soft X-ray harmonic radiation.** Having demonstrated the stability and reproducibility of the soft X-ray spectra, we further investigate the peculiar phase matching dependence at high pressure of long wavelength driven soft X-ray radiation by resorting to numerical simulations. A theoretical investigation is warranted, since the dependence of HHG at these high pressures is not readily investigated experimentally because the harmonic yield varies only moderately within the accessible pressure range, and it is stagnant with target position within the Rayleigh range, but it immediately drops below detection limits outside its phase matching region. Due to the high pressures required for long wavelength driven HHG, ionization is clearly a major determining factor for phase matching [29]. We therefore investigate the phase mismatch, analog to Ref. [22], for the full range of reachable target pressures as function of time in the laser pulse, and as function of position relative to the focal plane. We choose to illustrate the results of the investigation on the example of highest reachable photon energy of 500 eV in He. The identical behavior is found for cutoffs of 300 eV and 400 eV in He and for 300 eV and 375 eV in Ne – these figures are included in the Supplementary Information. Figure 3 displays the resulting phase matching maps, i.e. the on-axis phase mismatch $\Delta k(t,z)$, for the high energy end of the soft X-ray spectrum at 500 eV and for the full range of pressures in He. It is important to realize that the phase matching maps alone do not discriminate towards the ability to generate harmonics. Using the cutoff law for HHG, we can however determine the minimum pulse intensity needed to generate 500 eV photons which permits identification of the spatio-temporal region within the phase matching maps where the 500 eV photons are indeed generated. This region is indicated by the black encircled areas in Fig. 3, and the combined conditions evidence phase matched generation of 500 eV radiation. We find that, at low target pressures (Fig. 3a), phase matching is achieved for positions behind the focal plane (z > 0 mm), but is limited to a relatively short range of about 0.5 mm and a temporal region before the crest of the pulse. Increasing target pressure, we find a transition regime (Fig. 3b) in which the area of best phase matching moves to a region before the focus but still occurs temporally before reaching the crest of the pulse. Considering a range of photon energies, i.e. not only 500 eV, this regime indicates emission from multiple half-cycles and consequently emission of multiple attosecond bursts. The situation changes markedly at even higher target pressures (Fig. 3c) where optimal phase matching is temporally confined to the center of the pulse and for an interaction region symmetrically around the focal plane within the Rayleigh range (1 mm on either side of the focal plane). This region persists with even higher pressure (4 to 12 bar; see Supplement for a discussion on phase matching in focus), and continues narrowing temporally (Fig. 3d-f). We find a trend towards highest pressure (12 bar in Fig. 3f), which exhibits minimization of the temporal spread of the phase matched region and a symmetric distribution about the focal plane and within the phase matchable region. We note that the condition of minimal temporal spread and symmetric phase matching range coincides with the pressure (12 bar in He) for which we found the strongest harmonic signal experimentally (Fig. 1b). The transition from phase matching at low-pressure downstream from the focal plane to high pressure phase matching at, and around, the focal plane is explained by the interplay of the various phase matching terms. At low pressure, the interplay between dipole and geometric (Gouy) phase dominate phase matching while the neutral dispersion term is constant as a function of position and electron dispersion is negligible. This results in the well known conditions for best phase matching downstream from the focal plane [22]. In contrast, at high gas pressure, dispersion from electrons is dominant (see Supplement and Fig. S2) condequently resulting in a symmetric phase matching region around the focal plane because of the spatially symmetric electron dispersion. Repeating the identical analysis for He and for photon energies of 300 eV and 400 eV, we find the same behavior; see Figs. S3 and S4 (Figs. S5 and S6 for Ne). We also find from our simulation for a pressure of 6 bar in He, that the narrow temporal range coincides nearly identically for the vast range of photon energies of 300 eV, 400 eV and 500 eV (see Supplementary Fig. S7). These findings confirm the existence of a very narrow temporal phase matching window which provides the condition for emission of a single attosecond burst of soft X-ray radiation corresponding to the measured soft X-ray continuum. The lack of interference in the spectral measurements [25] shown in Fig. 2 further substantiates our conclusion. We note that the existence of such a window for high pressure phase matching was recently discussed [29], but was not proven due the lack of CEP control which is known to result in indistinguishability of a attosecond pulse continuum over random spectral shifts [30]. Moreover, the absence of CEP stability would result in random temporal shifts about the cycle duration (~6 fs for a 2000 nm pulse) which would inhibit attosecond resolution pump probe measurements. Here, we provide the first proof of these novel phase matching conditions through the demonstrated persistence and repeatability of the soft X-ray continuum and for varying CEP values (Fig. 2).

Our intuitive analysis suggests phase matching conditions resulting in temporal filtering of the macroscopic high harmonic signal and consequently in the reproducible emission of a single attosecond burst. Our findings are in excellent agreement with the experimentally observed parameters and it is this condition that we exploit to generate the high flux soft X-ray continuum shown in Fig. 2.

In conclusion, we have demonstrated the exploitation of a new phase matching regime which has allowed us to generate the first reproducible and CEP stable attosecond soft X-ray continuum covering the entire water window from 200 eV to 550 eV. We identify different and unexpected behavior of harmonic phase matching at the long wavelength driven high pressure regime compared to the commonly known conditions at 800 nm and discuss its implications on the stability and reproducibility of attosecond pulse generation. Having identified the crucial importance of CEP control we use the novel conditions to validate our findings by experimentally demonstrating unprecedentedly high flux of CEP controlled and repeatable attosecond continua covering the water window entirely. Our theoretical investigation indicates that the detected continuum is the result of a transient phase matching mechanism resulting in the generation of isolated broadband attosecond bursts.The simultaneous coverage of several absorption edges in combination with attosecond time resolution now enables following excitations and charge migration from one element to another. Thus we have bridged the gap between ultrafast time resolution and element specific probing. These results present a significant step towards studying the entire dynamics of the building blocks of biological life, within organic semiconductors, light harvesting devices and for molecular electronics.

# FIGURES

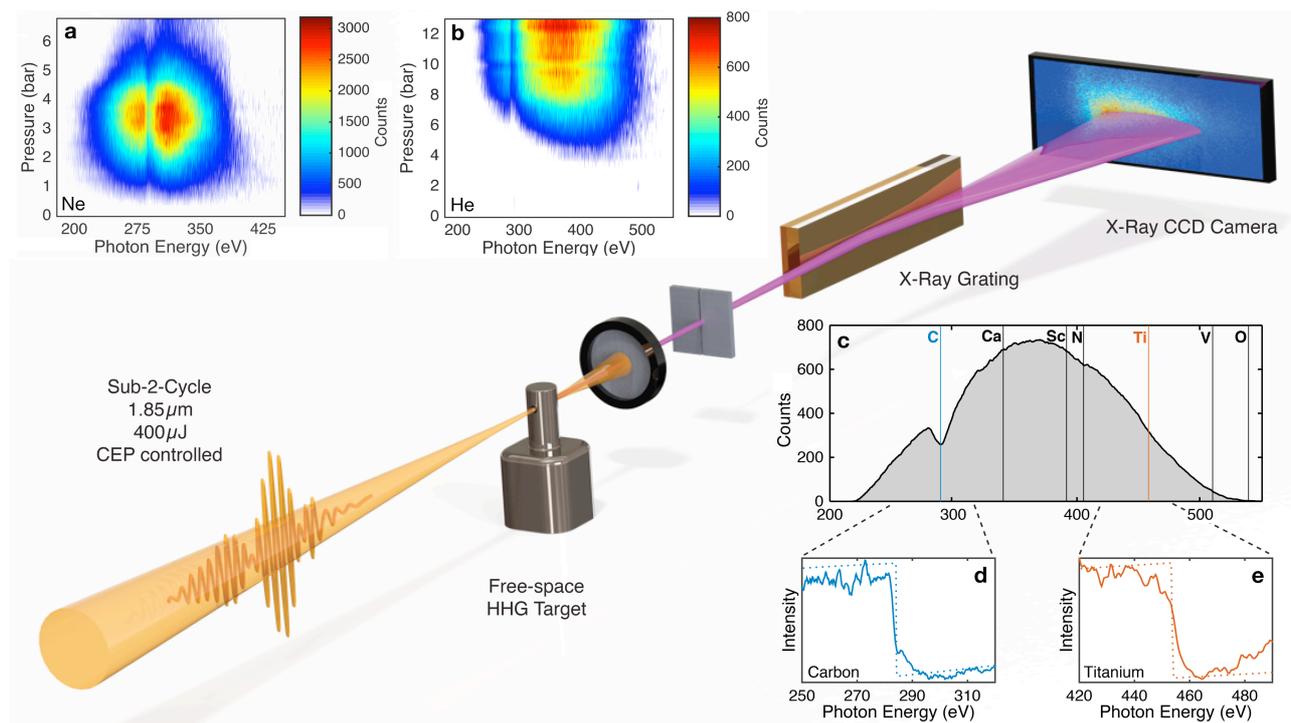

**Figure 1: Experimental setup, pressure dependence and spectral coverage.** Top: Pressure scans in Ne (a) and He (b). Clearly visible is the C K-shell absorption edge due to hydrocarbon residue in the beamline. Middle: Experimental setup consisting of a high pressure effusive target, a free-standing IR filter, and the home built spectrograph that consists of a 2400 lines per mm gold coated grating and a cooled X-ray CCD. Bottom right: (c) shows a spectrum from HHG in He (2 min integration time) with the reachable K-shell absorption edges indicated by solid vertical lines and L-shell absorption edges indicated by dashed vertical lines. The two graphs below show absorption measurements using foils of 200 nm of carbon (d) and titanium (e), where the K-edge at 284 eV and $L_{2,3}$-edges at 456 eV, are clearly evident.

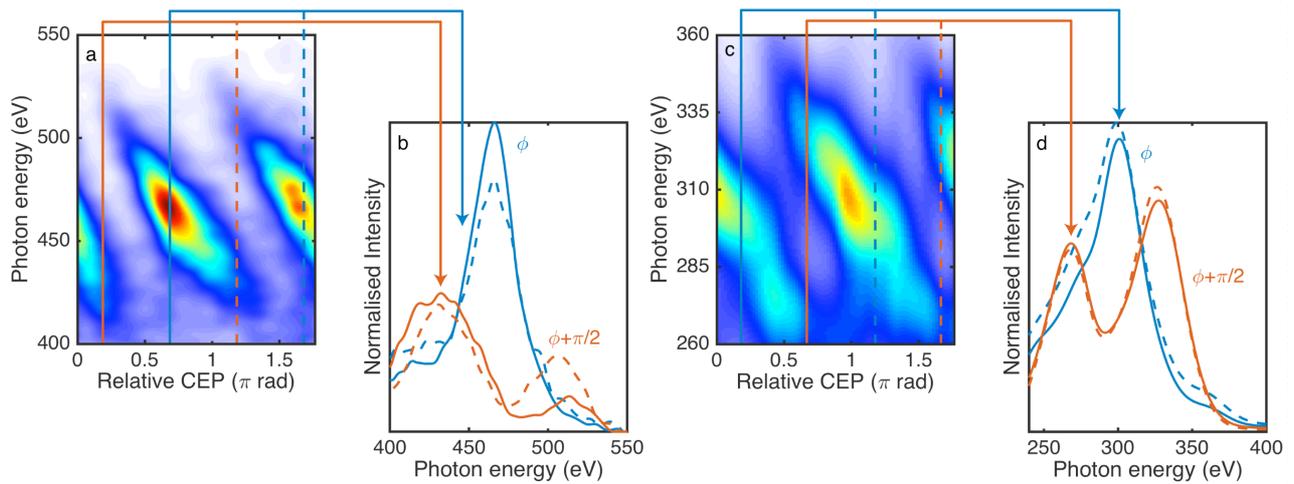

**Figure 2: CEP controlled soft X-ray emission and its spectral influence.** Shown in (a) and (c) are spectra generated from helium (6 bar) and neon (3.5 bar), respectively, which were acquired for varying CEP in steps of 90 mrad and with 30 s integration time each. The solid lines in (b) and (d) show the dramatic change of the spectral amplitude for two different CEP values; these values are indicated by the colored vertical lines in (a) and (c). The excellent CEP stability of the system results in clearly resolved spectra which repeat - as expected - with π rad periodicity.  This is shown by the dashed lines which are acquired for a π rad CEP offset compared with the matching solid colored lines – see (b) and (d).

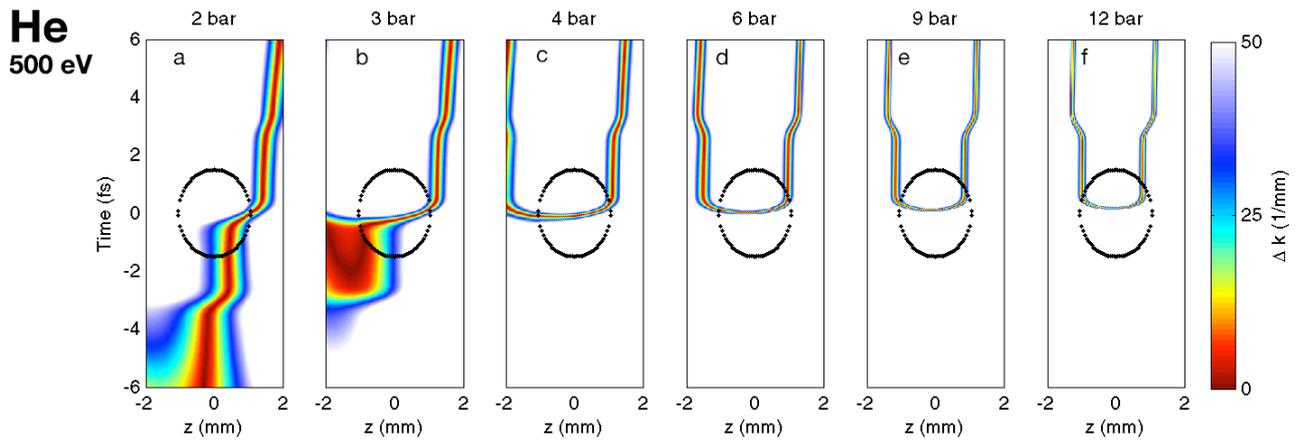

**Figure 3: Spatio-temporal phase matching maps as function of target pressure in He (500 eV).** Calculated on-axis phase mismatch as a function of propagation position and time within the pulse for 500 eV radiation generated in helium for our experimental conditions and target pressures of 2 to 12 bar (data for neon can be found in the Supplementary Information). Dark red indicates good phase matching and the back dotted oval area encloses the z-t space in which the field strength is sufficient to generate 500 eV radiation. At low pressure, phase matching occurs across the entire z-t range. At high pressure, good phase matching occurs only transiently within a narrow temporal window but across the entire Rayleigh length (here similar to the encircled spatial range).

# METHODS SECTION

**Experiment.** High-energy pulses from a Ti:Sapphire amplifier system with 40 fs duration, 7 mJ energy at 1 kHz repetition rate are frequency converted to 1.85 µm using an optical parametric amplifier [31], resulting in CEP-stable 0.8 mJ pulses with 40 fs duration. These pulses are spectrally broadened through nonlinear propagation in a hollow core fibre filled with 1.5 bar of argon, which both broadens the spectrum and introduces adequate dispersion for subsequent compression down to 12 fs with bulk material [18]. The CEP-stable pulses (see Supplementary Information) are then focused with an f=100 mm spherical mirror into a gas target, which has a length of 1.5 mm and consists of 0.3 mm diameter entrance and exit holes. The residual fundamental radiation is filtered with a 100 nm free standing aluminium foil and the transmitted radiation is refocused using a grazing-incidence ellipsoidal mirror (Zeiss). A homebuilt spectrograph, consisting of a motorised slit, flat-field imaging reflective grating (Hitachi, 2400 lines/mm) and a cooled, back-illuminated CCD (Princeton Instruments) is used to measure the spatio-spectral profile of the radiation in function of CEP. Photon counts are extracted using the measured spectra, the camera quantum efficiency and the measured grating diffraction efficiency.

**Simulations.** The phase matching calculations presented in Fig. 3 follow Ref. [22]. The on-axis phase mismatch between the source and the propagating X-rays is determined for a photon energy of 500 eV considering dispersion of both wavelengths from neutral gas and free electrons, the geometric phase (Gouy phase) of the fundamental as well as the dipole phase of the short trajectories. We do not consider long trajectories since they are effectively suppressed during propagation due to their non-collinear off-axis emission and filtering on pumping apertures of our beamline. We use a Gaussian focus with a waist of 54 µm for the calculations and find the time-dependent fraction of free electrons for a peak intensity of 0.49 PW/cm$^2$.

**References**

1. Krausz, F. & Ivanov, M. Y. Attosecond physics. *Rev. Mod. Phys.* **81**, 163-23497 (2009).

2. Belshaw, L. et al. Observation of Ultrafast Charge Migration in an Amino Acid. *Phys. Chem. Lett.* 3, 3751-3754 (2012).

3. Schiffrin, A. et al. Optical-field-induced current in dielectrics. *Nature* **493**, 70-74 (2013).

4. Kling, M. F. et al. Control of electron localization in molecular dissociation. *Science* **312**, 246-248 (2006).

5. Sansone, G. et al. Electron localization following attosecond molecular photoionization. *Nature* **465**, 763-766 (2010).

6. Pfeifer, T., Spielmann, C. & Gerber, G. Femtosecond X-ray science. *Rep. Prog. Phys.* **69**, 443-505 (2006).

7. Föhlisch, A. et al. Direct observation of electron dynamics in the attosecond domain. *Nature* **436**, 373-376 (2005).

8. Schnadt, J. et al. Experimental evidence for sub-3-fs charge transfer from an aromatic adsorbate to a semiconductor. *Nature* **418**, 620-623 (2002).

9. Cavalleri, A. et al. Band-Selective Measurements of Electron Dynamics in $VO_2$ Using Femtosecond Near-Edge X-Ray Absorption. *Phys. Rev. Lett.* **95**, 067405 (2005).

10. Hentschel, M. et al. Attosecond metrology. *Nature* **4147**, 509-513 (2001).

11. Sola, I. J. et al. Controlling attosecond electron dynamics by phase-stabilized polarization gating. *Nat. Phys.* **2**, 319-322 (2006).

12. Kienberger, R. et al. Atomic transient recorder. *Nature* **427**, 817-821 (2004).

13. Sansone, G. et al. Isolated single-cycle attosecond pulses. *Science* **314**, 443-634 (2006).

14. Goulielmakis, E. et al. Single-cycle nonlinear optics. *Science* **320**, 1614-1674 (2008).

15. Colosimo, P. et al. Scaling strong-field interactions towards the classical limit. *Nat. Phys.* **4**, 386-389 (2008).

16. Popmintchev, T. et al. Extended phase matching of high harmonics driven by mid-infrared light. *Opt. Lett.* **33**, 2128 (2008).

17. Xiong, H. et al. Generation of a coherent x ray in the water window region at 1 kHz repetition rate using a mid-infrared pump source. *Opt. Lett.* **34**, 1747-1749 (2009).

18. Ishii, N. et al. Carrier-envelope phase-dependent high harmonic generation in the water window using few-cycle infrared pulses. *Nature Commun.* **5**, 3331 (2014).

19. Cousin, S.L. et al. High-flux table-top soft x-ray source driven by sub-2-cycle, CEP stable, 1.85-µm 1-kHz pulses for carbon K-edge spectroscopy. *Opt. Lett.* **39**, 5383-5386 (2014).

20. Silva, F. et al. Spatiotemporal isolation of attosecond soft X-ray pulses in the water window. *Naure. Commun.* **6**, 6611 (2015).

21. Balcou, P. & L'Huillier, A. Phase-matching effects in strong-field harmonic generation. *Phys. Rev. A* **47**, 1447-1459 (1993)

22. Balcou, P. et al. Generalized phase-matching conditions for high harmonics: the role of field-gradient forces. *Phys. Rev. A* **55**, 3204–3210 (1997).

23. Kazamias, S. et al. Pressure-induced phase matching in high-order harmonic generation *Phys. Rev. A* **83**, 063405 (2011).

24. Chen, M.-C. et al. Bright, Coherent, Ultrafast Soft X-Ray Harmonics Spanning the Water Window from a Tabletop Light Source. *Phys. Rev. Lett.* **105**, 173901 (2010).

25. Haworth, C. A. et al. Half-cycle cutoffs in harmonic spectra and robust carrier-envelope phase retrieval. *Nature Phys.* **3**, 52-57 (2007).

26. Teichmann, S. M. et al. Importance of intensity-to-phase coupling for water-window high-order-harmonic generation with few-cycle pulses. *Phys. Rev. A* **91**, 063817 (2015).

27. Mashiko, H. et al. Double Optical Gating of High-Order Harmonic Generation with Carrier-Envelope Phase Stabilized Lasers. *Phys. Rev. Lett.* **100**, 103906 (2008).

28. Ferrari, F. et al. High-energy isolated attosecond pulses generated by above-saturation few-cycle fields. *Nature Phot.* **4**, 875-879 (2010).

29. Chen, M.-C. et al. Generation of bright isolated attosecond soft X-ray pulses driven by multi-cycle mid-infrared lasers. *Proc. Nat. Acad. Sci. U.S.A.* **111**, E2361-E2367 (2014).

30. Baltuska, A. et al. Attosecond control of electronic processes by intense light fields. *Nature* **421**, 611-615 (2003).

31. Silva, F. et al. High-average-power, carrier-envelope phase-stable, few-cycle pulses at 2.1 µm from a collinear BiB$_3$O$_6$ optical parametric amplifier. *Opt. Lett.* **37**, 933-935 (2012).

**Acknowledgements**

We acknowledge financial support from the Spanish Ministry of Economy and Competitiveness, through the "Severo Ochoa" Programme for Centres of Excellence in R&D (SEV-2015-0522), FIS2014-56774-R, FIS2014-51478-ERC and the Catalan Agencia de Gestió d'Ajuts Universitaris i de Recerca (AGAUR) with SGR 2014-2016. This research has been supported by Fundació Cellex Barcelona, the Fundação para a Ciência e a Tecnologia (grant SFRH/BD/69913/2010), the European Union's Horizon 2020 research and innovation programme under the Marie Sklodowska-Curie grant agreement No. 641272 and Laserlab-Europe (EU-H2020 654148).

**Author Contributions**

J.B. conceived the experiment and designed the attosecond beamline. M.H., S.L.C. and J.B. developed the laser source. S.M.T., F.S. and S.L.C. performed the experiment. The manuscript was written by S.M.T. and J.B.